\documentclass[twocolumn,showpacs,pre]{revtex4}
\usepackage{graphicx}
\newcommand{\myfig}[2] {
\begin{figure}[h]
\centerline{ \includegraphics[width=8cm]{#1.eps} }
\caption{#2} \label{#1}
\end{figure} }

\newcommand{\myfigsmall}[2] {
\begin{figure}
\centerline{ \includegraphics[width=6cm]{#1.eps} }
\caption{#2} \label{#1}
\end{figure} }

\begin{document}

\title{\bf
Branching mechanism of
intergranular crack propagation
in three dimensions
}
\author{M.~Itakura}
\author{H.~Kaburaki}
\affiliation{Center for Promotion of Computational Science and Engineering,
Japan Atomic Energy Research Institute,
Taito-ku, Higashiueno 6-9-3, Tokyo 110-0015, Japan}
\author{C.~Arakawa}
\affiliation{
Interfaculty Initiative in Information Studies,
University of Tokyo,
7-3-1 Hongo, Bunkyo-ku, Tokyo 113, Japan
}
\date{\today}

\begin{abstract}
We investigate the process of slow
intergranular crack propagation
by the finite element method model and show that branching
is induced by partial arresting of a crack front
owing to the geometrical randomness of grain boundaries.
A possible scenario for the branching instability of
crack propagation in  a disordered continuous medium is also discussed.
\end{abstract}
\maketitle

%%\section{intro}
The morphology of cracks has been the subject of
intensive studies in recent years.
Experimental observations of the universal roughness exponent
$\eta \sim 0.8$
of the fracture surface \cite{eta-exp1, eta-exp2}
have been stimulating theoretical and numerical studies
of relevant models.
%,and similar exponent has been reproduced 
%in numerical simulation \cite{eta-mc}.
Another interesting subject is the branching behavior
of fast-propagating cracks: There seems to be a dynamic
branching instability that is common in various kinds of
amorphous materials \cite{branch-exp}, and this branching
instability has been numerically reproduced \cite{branch-phf, branch-mc2}.
In the brittle fracture of gels,
a different kind of branching has been observed \cite{gel}.

%E.Bouchaud, J.Phys. Condens. Matter 9, 4319(1997)
%Branch theory   PRL 65 1784 (1990) R.C.Ball R.Blumenfeld
%MD-like FEM, phase field

Branching is also observed in slowly propagating
cracks such as intergranular stress corrosion cracking (IGSCC),
which occurs when a polycrystalline metal or alloy is subjected to
both tensile stress and a corrosive environment (See Fig. \ref{f0schem}),
such as
%%atmosphere of swimming pool (chlorine) or 
nuclear reactor coolant
(irradiated water). The corrosive agent
selectively corrodes the grain boundary (GB) near the crack tip,
which is under tensile
stress, and the crack propagates along the GBs
exhibiting  typical branching patterns.
Empirical
relations between the mode-I stress intensity factor $K_I$ at the crack tip
and crack propagation
velocity $v$ is used to assess the safety of structural materials,
and they usually take power-law form  $v = C (K_I-K_{Ic})^a$,
%%% NUREG-0313, Rev.2 WS Hazelton and WH Koo
where $C$ and $a$ are parameters that depend both on material and environment,
and $K_{Ic}$ is a critical value of $K_I$ at which 
the crack begins to propagate.
The typical velocity of crack propagation
of IGSCC under an industrial environment is of the order 0.1 to 1  mm per year.
%Although the exact mechanism of SCC and
%is still unclear, one can at least say that
%the stronger the tensile stress, the faster crack propagates.

\myfigsmall{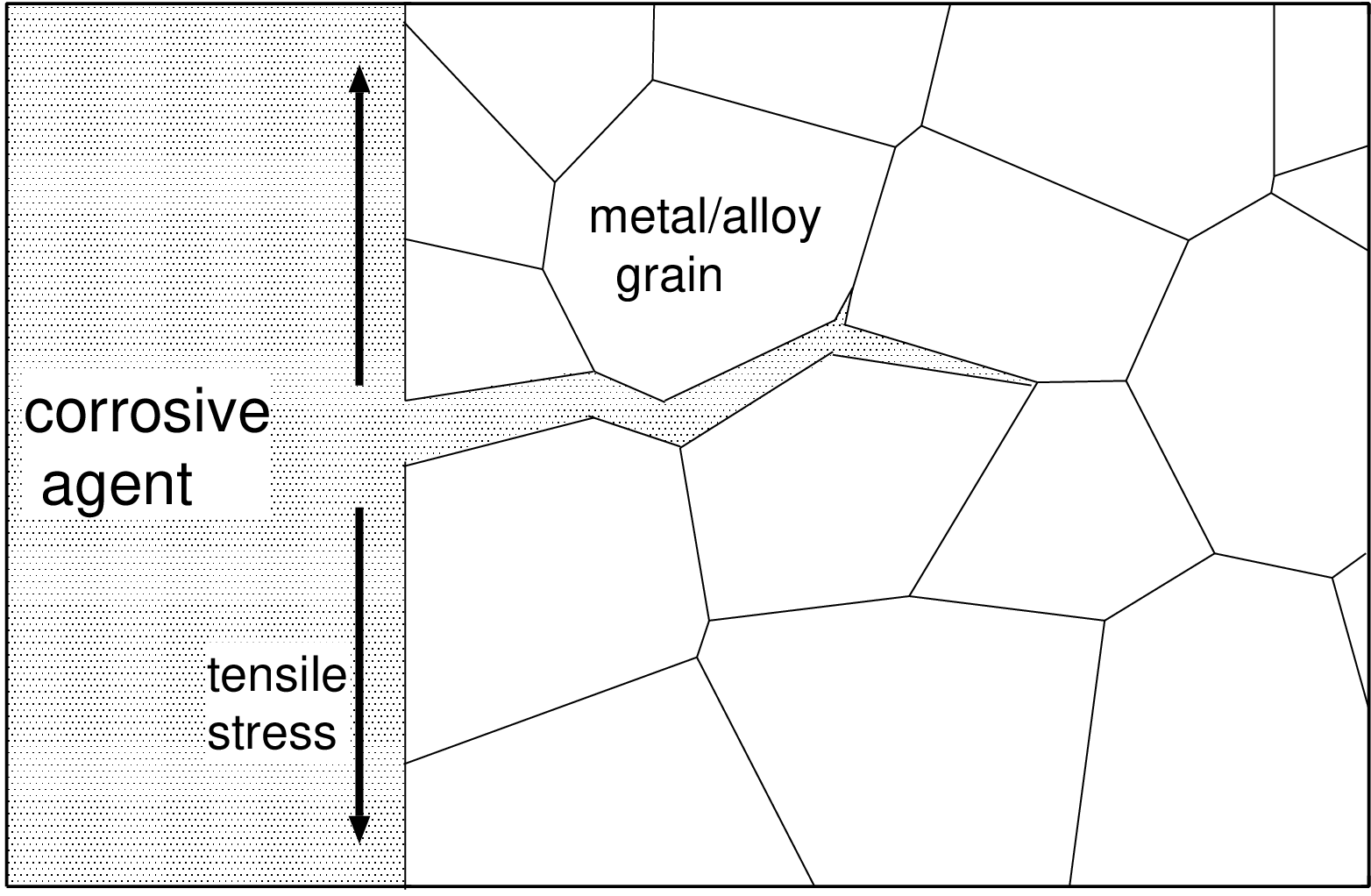}{Schematic depiction of intergranular
stress corrosion cracking. A corrosive agent diffuses through
the opened crack and corrodes grain boundaries under tensile stress.}

Naively, the branching of intergranular cracking may seem 
obvious because there are numerous GB triple junctions
where 
a crack front has a chance to branch, 
but in reality, it is not so simple:
If a branch occurred at a triple junction,
the stress concentrates on the longer branch and thus 
enhances its propagation, screening the stress of shorter branch,
eventually suppressing its propagation before it grows to a 
length compatible with the GB length.
In the present paper, we model the intergranular crack 
propagation process and carry out
numerical simulations,  and show that the branching occurs
even when the explicit branching at GB triple junctions is forbidden.

%%%%%%%%%%%% \section{The model}
%We simulate intergranular crack propagation using 
%finite element method (FEM) like model of polycrystal,
%in which cracks proceed along the GBs.
%We simulate intergranular crack propagation using
At the first stage of the simulation, polycrystalline GBs are 
prepared using random Voronoi tessellation \cite{voro} of a cube of 
dimensionless size $1.0\times 1.0 \times 1.0$,
and a crack is assumed to propagate through these GBs.
%A finite element method (FEM) like model of polycrystal,
%in which cracks proceed along the GBs.
In this paper, 12 000 grains are used.
Tensile stress along the $y$ axis (see Fig. \ref{f1sysgeom}) is applied
as constant loads on the $y=0$ and $y=1$ plane of the cube,
and local stress distribution (by which the crack is driven)
is calculated using a simplified finite-element-method (FEM)  model.
The FEM nodes are placed at vertices of grains, centers of
GB surfaces, and centers of grains. Each grain
is decomposed into tetrahedral FEM elements,
which contain two vertex nodes,
one surface center node, and one grain center node (Fig. \ref{f2fem}(a)).
To glue the grains together,
a very thin FEM element is placed at each GB,
which is made up of six-node triangular elements (Fig. \ref{f2fem}(b)).
The thickness of this gluing element is set to $10^{-4}$.
When a GB fails, the elastic constants of the corresponding
gluing element are set to zero.
To enhance the stress concentration at the crack tip,
the initial crack is prepared
by separating all the GBs between two grains
whose grain centers are above and below a plane $y=0.5$
and lie in a region $z<0.2$.
The crack then proceeds in the $z$ direction.

\myfigsmall{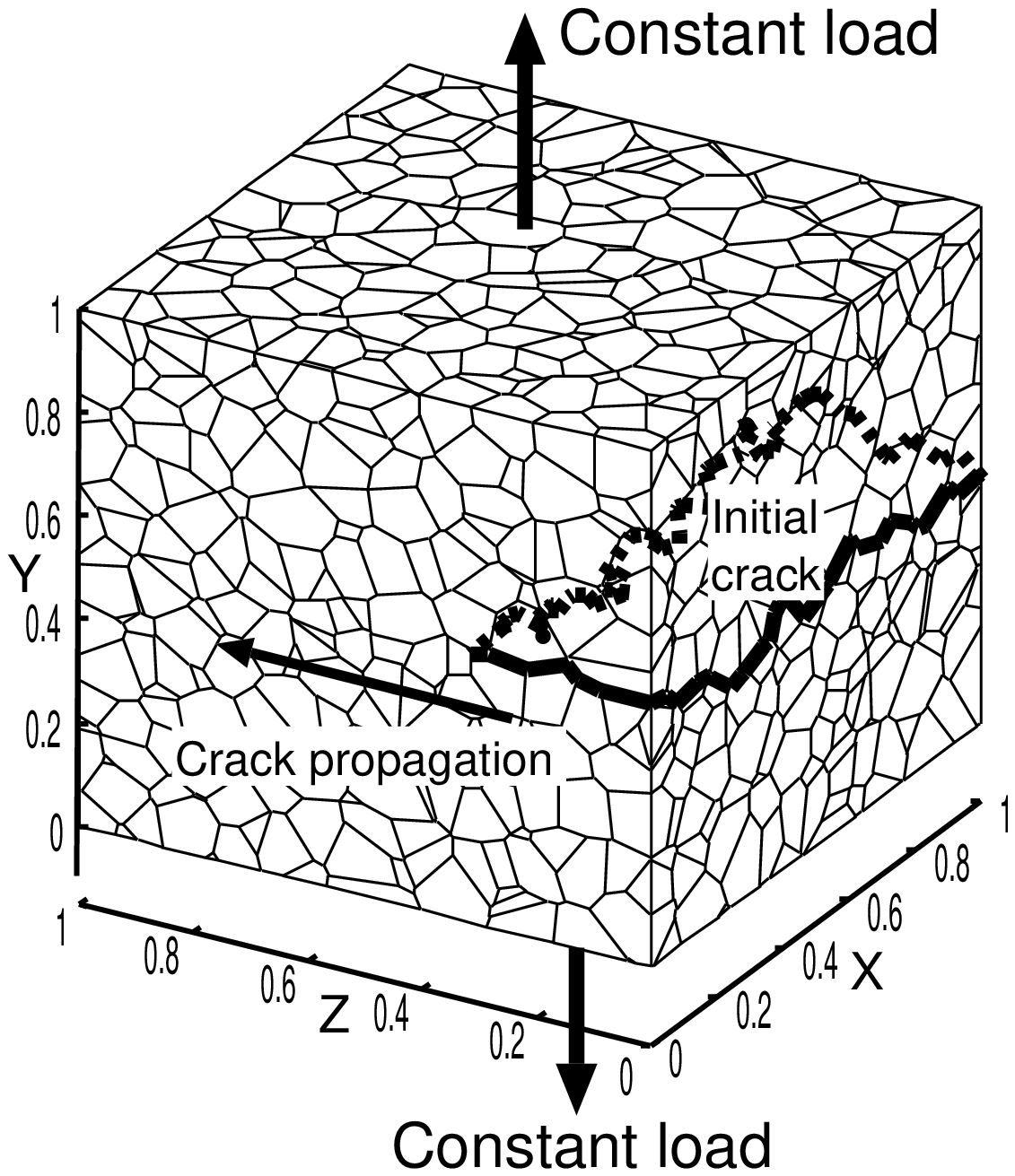}{Geometry of a simulation cell.
An initial crack is placed at a region $y\sim 0.5$, $z < 0.2$ and
constant load is imposed on the upper ($y=1$) and lower ($y=0$)
surface of the cube.}

\myfigsmall{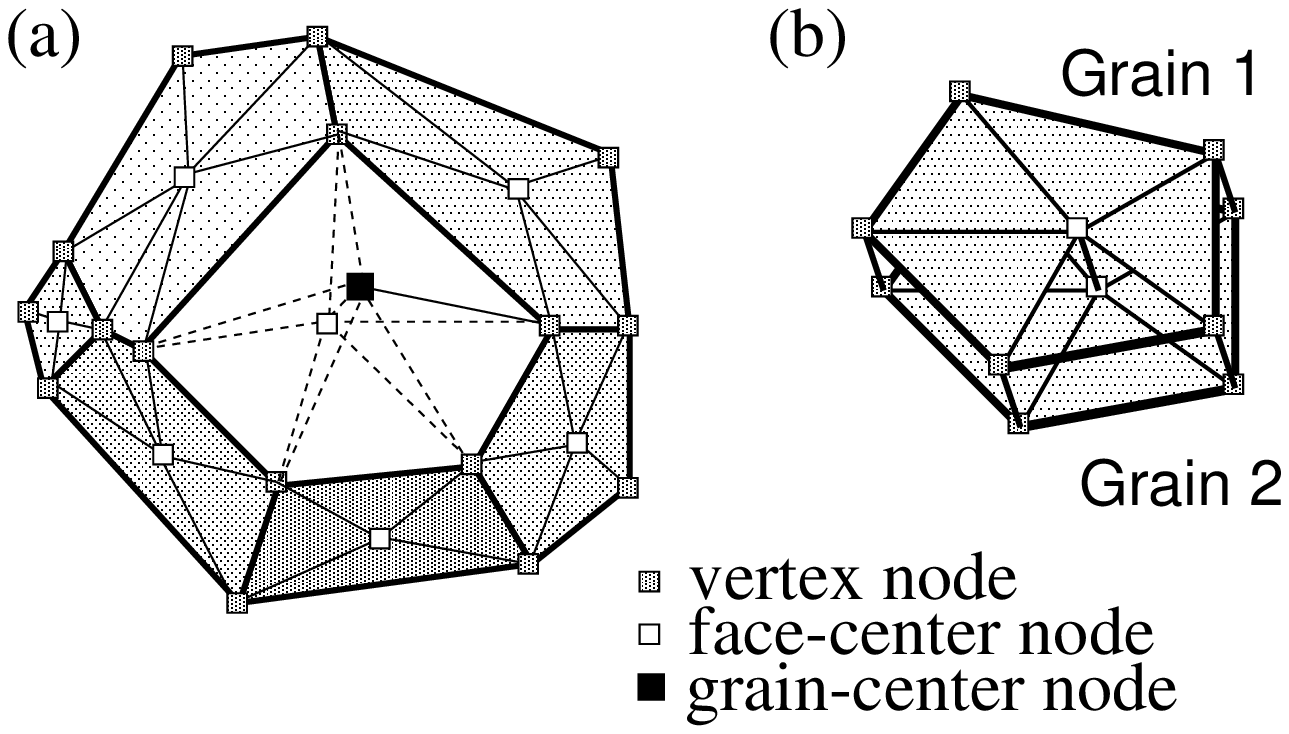}{Construction of FEM elements.
(a) Nodes are placed at vertex, face center, and body center
of each grain. (b) Each grain is glued together by thin triangular
elements.}

The most crucial and difficult part of this kind of modeling studies
is a
determination of the crack propagation rule.
In this paper,
a certain GB is selected based on the stress distribution,
and is separated completely.
Therefore the GBs fail one at a time,
which is similar to the rule employed in the so-called network models,
such as random fuse models \cite{fuse} and random 
spring models \cite{spring}.
Although the validity of this rule for the simulation of
intergranular cracking is quite unclear
because the nearby stress distribution may change significantly
while the crack front proceeds through the selected GB
and may initiate another GB failure,
there are two reasons why we employ this rule.
Firstly, to track the continuous propagation of the crack front
along a GB requires very fine FEM meshing around the
crack tip, or alternatively, the FEM mesh must be reorganized around
the crack tip each time the crack proceeds by a small amount.
Both methods require extensive computational power and a fairly complex
simulation code, and the number of grains will be
severely restricted.
Secondly, it is plausible to assume
 that the crack front is arrested at the triple
junctions of GBs for a long time; thus the propagation
process may be treated as a series of discrete events of GB
failure.

The rule to determine which GB to separate is that
we choose a GB on which the strongest tensile
stress normal to its surface is imposed.
Considering that the stress diverges near the crack tip
as $K_I r^{-1/2}$ in the linear elastic theory, where 
$r$ is a distance to the crack tip
and $K_I$ is a mode-I stress intensity factor,
%which depends on the material, crack shape, and external stress.
we choose among the GBs that are adjacent to the
crack tip. 
In the present model, the crack tip is defined as a set of GB triple
junctions that is shared by one fractured and two unfractured GBs.
This restriction forbids explicit branching
at triple junctions, as well as isolated crack initiation 
in the unfractured interior, and  distinguishes 
the present model from the network models that are mainly used
to study crack morphology.
After the selected GB fails, the elastic matrix is updated
and the fracture process is repeated.

Local stress is calculated by
standard linear elastic theory \cite{landau}.
Since only the linear theory is used in the present paper,
one can arbitrarily scale the stress and strain.
We set Young's modulus $E$ to unity and assume the elastic
properties to be isotropic; therefore the only elastic parameter
to be considered is Poisson's ratio $\nu$, and simulations are carried out
for several values of $\nu$.
%The FEM nodes are placed at vertices of grains, centers of
%GB surfaces, and centers of grains. Each grain
%is decomposed into tetrahedral FEM elements,
%which contain two vertex nodes,
%one surface center node, and one grain center node (Fig. \ref{f2fem}a).
%To glue the grains together,
%a very thin FEM element is placed at each GB,
%which is made up of six-nodes triangular elements (Fig. \ref{f2fem}b).
%The thickness of this gluing element is set to $10^{-4}$.
%When a GB fails, elastic constants of the corresponding
%gluing element are set to zero.
To evaluate the tensile stress acting on a GB, a normal
component of a difference of
displacements between two nodes that lie on both sides of the gluing
element is calculated at each vertex of a GB.
Here we only investigate vertices on the crack
tip, where maximum of tensile stress occurs.
The FEM mesh we use is very coarse compared to
engineering studies, in which progressively finer mesh 
is used around the crack tip.
However, the main concern in the engineering studies
is to evaluate precisely a stress intensity factor
at the crack tip and to determine whether it is
greater than the critical value above which the crack
propagates catastrophically. Therefore,
in these studies, the initial crack is usually assumed to be a
semicircular microcrack and the propagation process
is not studied.

There have been several numerical studies of 
intergranular crack propagation in which
some simplifying approximations are  used,
such as redistributing the stress of a failed
surface equally to the neighbor surfaces \cite{ig-mf}.
%%restriction of crack front shape to solid-on-solid type
%%zig zag lines\cite{ig-sos}, and so on.
The present paper
% is the first work which 
simulates and evaluates intergranular
cracking process with full  geometrical modeling of three-dimensional
grain boundaries
and evaluates an approximate local stress field (if crude) by which
crack is driven.
A system of linear equations has been solved using the
BiCG-stab (bi-conjugate gradient stabilized)
iterative solver. The dimension of the vector was
about 1 500 000, and the number of nonzero elements of the
symmetric elastic
matrix was about 28 000 000 $\times 2$ in the case of 12 000 grains.
Most of the CPU time of the simulation was spent in the
solver routine, which has been vectorized and run on a NEC SX-6
vector processor. The overall CPU time needed to
carry out 1 500 steps of GB fracture was about 6 h.

%%\section{simulation result}
Figure \ref{f3result} shows fracture surface projected onto 
the $X-Z$ plane,
obtained from the simulation of the $\nu=0.25$ case:
the black area shows the fracture surface,
and the light gray areas show the branched fracture surfaces.
In this case, 
more than 20 percent of the fracture surface area
(projected onto the $X-Z$ plane) is covered
by the branched surface, even though explicit branching at
GB triple junctions is forbidden.
Figure \ref{f4covrat} shows convergence of the ratio of 
the branch surface
plotted against the projected fracture surface area
for the three typical cases of $\nu=0.0$ (spongy),
$\nu=0.25$ (a typical value of metals), and $\nu=0.49$ (rubbery).
It can be seen that
in each case the ratio converges to a value $0.2$--$0.3$, and the
branching behavior does not vary drastically, even in the extreme
cases of $\nu=0.0$ and $\nu=0.49$.

\myfig{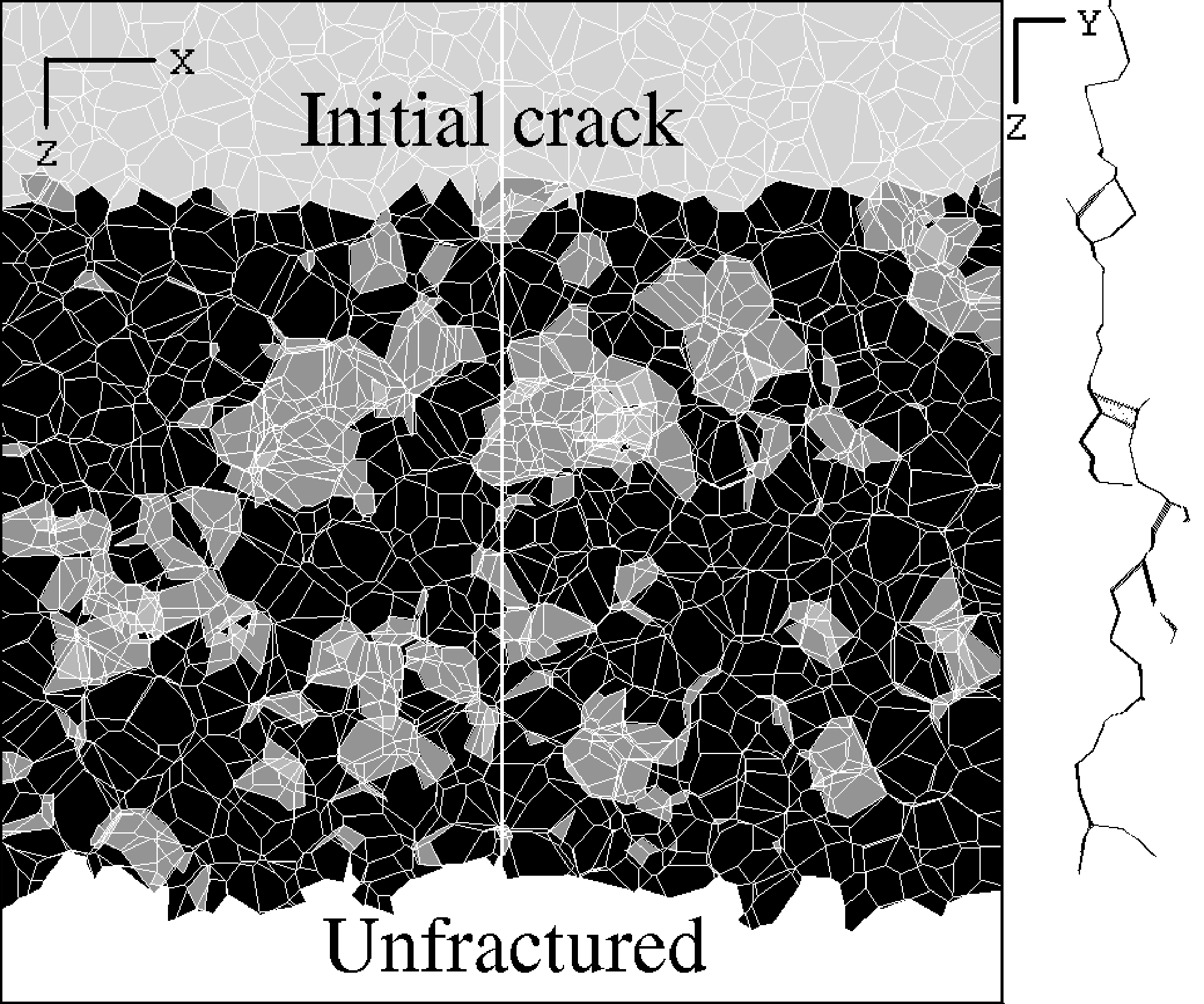}{Fractured surface observed in the simulation
of $\nu=0.2$ case, projected onto the $X-Z$ plane.
The black and gray areas show the fracture surface and branch fracture
surface, respectively. The cross section of the fracture surface 
in the center (bold white line) is shown on the right side.
}

\myfig{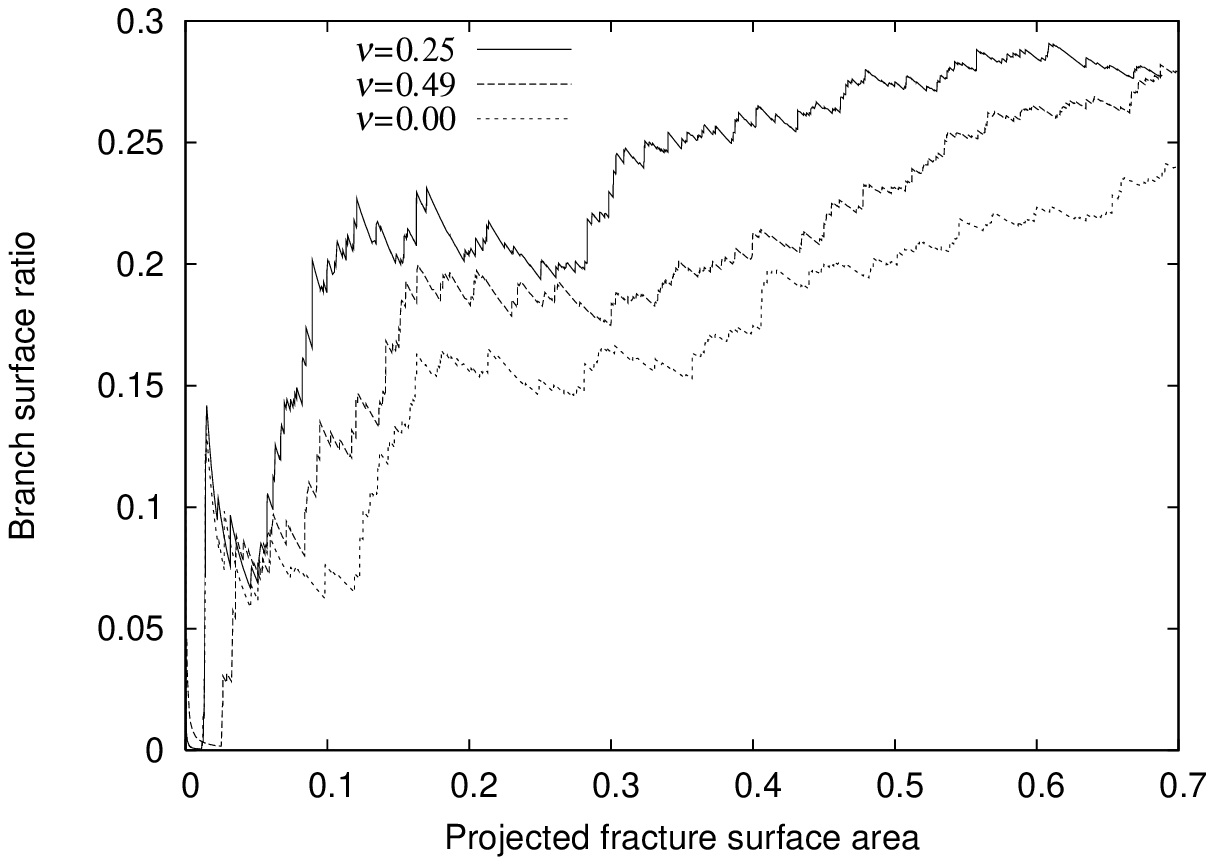}{
Ratios of branch surface area
plotted against the projected fracture surface area
for several values of Poisson's ratio $\nu$.
}

Figure \ref{f5schemlat} schematically depicts the typical branching
mechanism observed in the simulation:
(a) The crack front is arrested at the sloped surface $S0$, 
where mode-I stress is reduced by a factor of $\cos \theta$
where $\theta$ is the inclination angle of $S0$ out of the $X-Z$ plane.
(b) The crack initiates at the point $v$ and propagates along 
the much horizontal surface $S1$, and intersects %already 
the fracture surface at the segment $L0$.
After this kind of branch is formed, the branched
crack front circumvents
the arresting GB and continues propagation, and eventually merges
again to resume intact crack front line and leaves a branch behind
that consists of several GBs.
Although the branch length is only of an order of several GBs,
so frequent a branching, as observed in this simulation,
will significantly affect the crack propagation velocity if
we would construct a time-driven crack propagation rule
and simulate it.
As for the long-length scale properties of the fracture surface,
an estimate of roughness exponent is of interest,
but the obtained fracture surface,
which consists of about 1 500 GBs, is not
large enough to observe such quantities.

\myfig{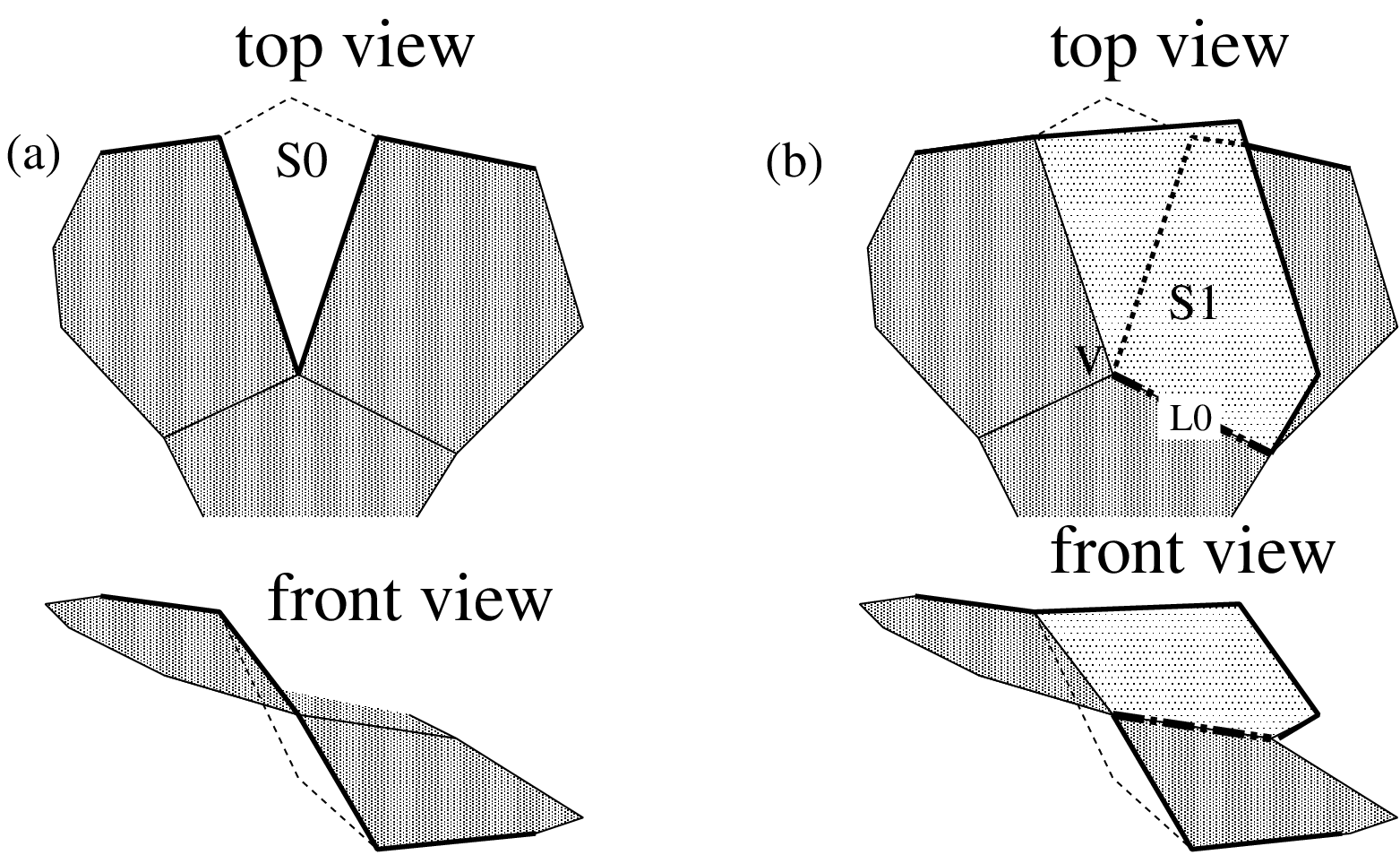}{Schematic depiction of the typical crack branching process
observed in the simulation: (a) just before the branching,
(b) just after the branching. See the main text for details.}

%depends on the material 
%granular /gel polymer  brittle/ductile
%roughness exponent
%based on percolation model
%theory:void coalescence

%%%%===
%local tensile stress perpendicular to the granular surface. 
%When the tensile stress is negative, the crack front does not proceed.
%These rules of our model are based on the experimental facts that
%the SCC does not occur under compressive stress,
%and the velocity of macroscopic crack propagation of SCC is proportional
%to some material dependent power of stress intensity factor.
%concentrates on the longer branch, enhancing the grows of faster one.
% and sub-branch will emanate from both ends of the segment to relax the stress.

%%%\section{continuum case}
So far, we have shown that in intergranular crack propagation,
branching frequently occurs owing to the partial arresting of
crack front. Here
%In this section, 
we infer that
this branching behavior
%similar to the one described above %in the previous section
may also occur in more general cases of
crack propagation in  a disordered continuous medium,
under some modest assumptions.
First, we assume that the crack propagation velocity $v$
%is a function of local stress intensity factor at the crack tip
%$$v = f(K_I, K_{II}, K_III),$$
%and that the function $f$ 
mainly and strongly depends on $K_I$, that is,
$$ \frac{\partial v}{\partial K_I} \gg
\frac{\partial v}{\partial K_{II}}
, \,\,\,
\frac{\partial v}{\partial K_I} \gg \frac{\partial v}{\partial K_{III}}
,\,\,\,
\frac{\partial v}{\partial K_I} \gg \frac{v}{K_I}.$$
For example, a power-law function
$v=\left[ K_I^2 + \epsilon (K_{II}^2 + K_{III}^2) -K_c^2 \right]^{a/2}$
satisfies these conditions when $\epsilon \ll 1$ and $a \gg 1$.
Secondly, local mode-II stress is assumed to change the crack
propagation direction out of the current crack plane so that
the mode-I stress normal to the plane increases.

Now consider a straight crack front propagating in an inhomogeneous
continuous medium. When the front crosses a small region where 
mode-II stress is locally induced by inhomogeneous
elastic properties, a ¡Èhump¡É along the vertical direction
is generated (Fig. \ref{f6conthump}).
This hump will be eventually lowered owing to the interactions 
between crack front segments, if the propagating velocity of
each crack front segment does not vary strongly.
But the sloped segment feels smaller mode-I stress (by a factor of 
$\cos \theta$,
where $\theta$ is an inclination angle of the segment) and its propagating
velocity becomes much smaller, say, by a factor of
$\cos^a \theta$. This effect may be compensated to some
degree, because mode-I stress concentrates on a segment lagged behind.
If this compensation is not sufficient, branching of the crack front
can occur through a mechanism described below and shown 
in Fig. \ref{f7contschem}:
%Although fluctuation along z direction is suppressed
%by a mechanism that stress concentrates to
%a segment which is lagged behind and is accelerated,
%there is no mechanism which suppress the fluctuation along 
%the y direction, since the mode II stress vanishes once the
%propagation direction returns to the z direction
%and sloped front feels only mode III stress which only affects 
%in-plane direction of crack propagation.
(a) The sloped section is lagged behind, owing to the
weak mode-I stress at the crack front
(b) The left and the right side of the segment bulge
inward.
(c) The bulged segments further proceed and eventually overlap
each other. Then one part shields the stress and continues
to proceed, while the other slows down.
(d) Owing to mode-II stress induced by the interaction between 
the crack front segments of overlapped parts,
each part gets closer and eventually intersects.
(e) Here a segment of the triple junction, or a root of a 
branch, is formed. 
(f) A branched ``tongue'' is left behind and the crack front
(now intact) proceeds further.
In this way, many small branches are left behind the sweeping crack front
also in continuum case.

\myfig{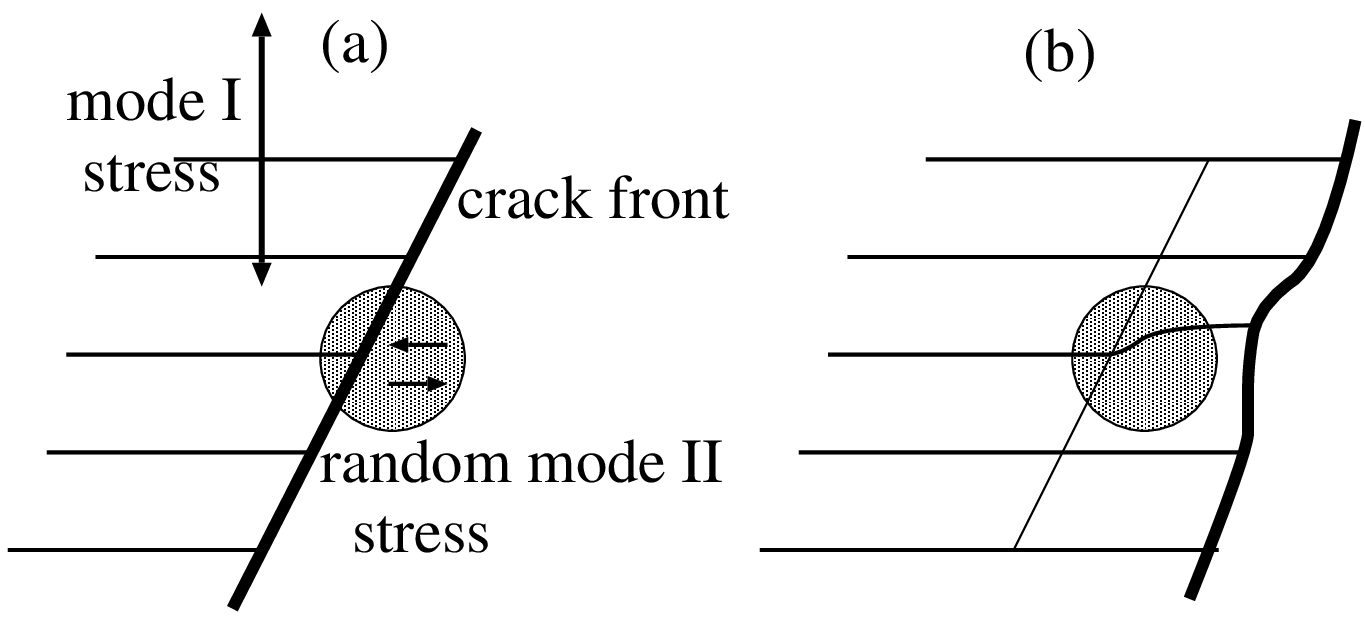}{Schematic picture of humping of crack front
propagating in an inhomogeneous medium}

%%%\section{discussion}
In summary,
we have modeled and simulated slow intergranular crack propagation,
and found that branching of a crack frequently occurs even if
explicit branching at grain boundary triple junctions is forbidden.
In real intergranular fracture,
random crystallographic anisotropy of the elasticity of
each grain produces
inhomogeneous stress distribution \cite{aniso} and  will
enhance crack arresting that leads to more frequent branching.
In addition, in the case of polycrystalline metals,
a certain portion of GBs are small-angle grain boundaries
that are very resistant to fracture and corrosion. Thus,
there are numerous arresting GBs and the branching may be
strongly enhanced, as observed in IGSCC.
We have also inferred that the crack branching mechanism observed 
in the simulation of the discrete model may occur in more general
cases of crack propagation in a disordered continuous medium,
under some modest assumptions on the relation between
crack propagation velocity and stress intensity factors.
A direct numerical simulation of this continuum case is expected.

The authors would like to thank Masayuki Kamaya,
Takashi Tsukada, and Yoshiyuki Kaji for helpful remarks.

\myfig{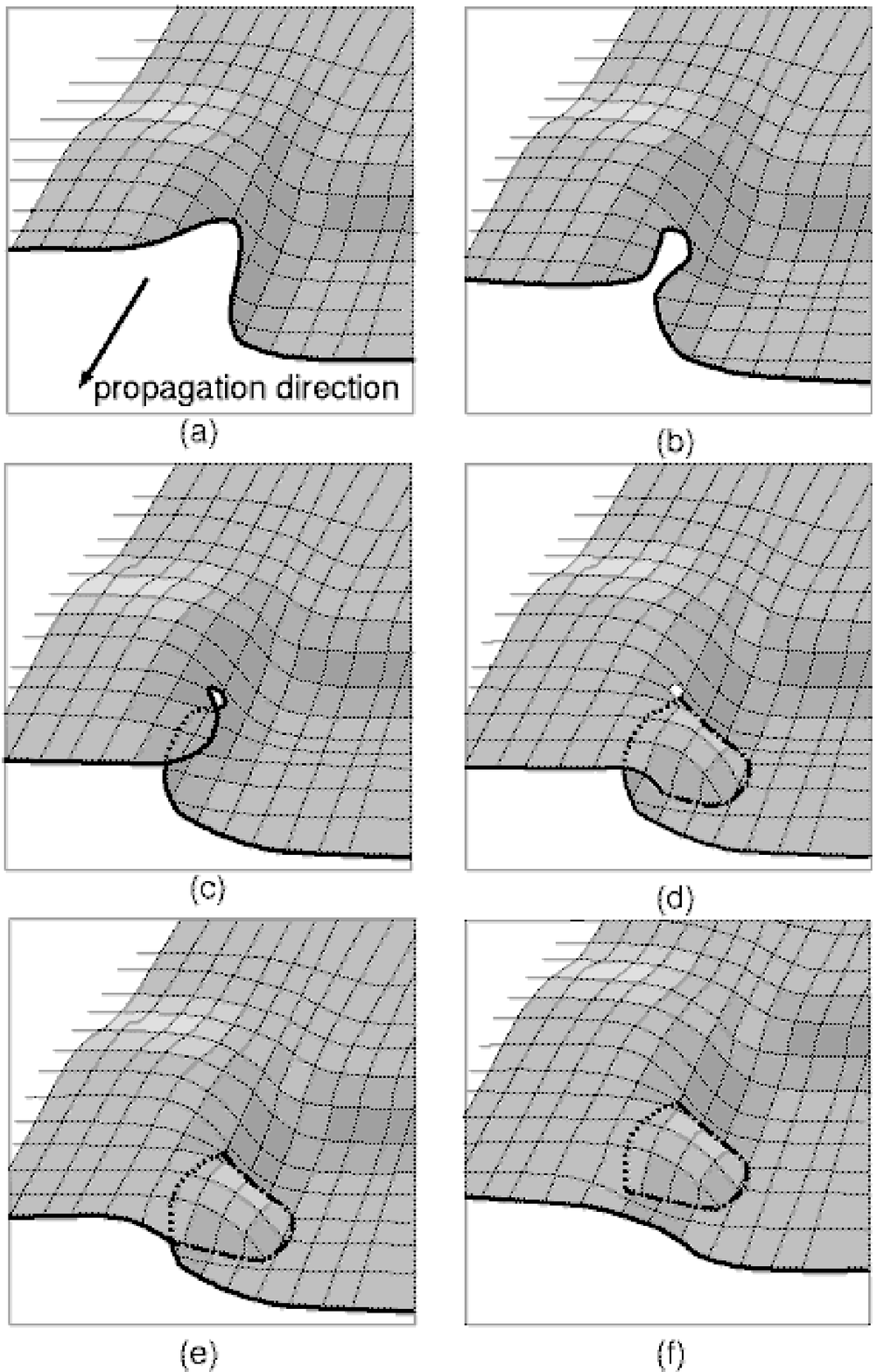}{Schematic picture of the crack branching process
expected in a disordered continuous medium.
Bold solid lines and bold dotted lines are the crack tip and
crack tip under the fracture surface, respectively.
The dashed line denotes the triple junction of fracture surface.
See the main text for details.
}

%%Although we concentrated on IGSCC, the mechanism of crack branching
%%seems to be general in crack propagation in disordered medium.
%if there is a region which strongly arrests the crack propagation,
%crack front 
%%In case of crack propagation in a random continuum media,


\begin{thebibliography}{99}

\bibitem{eta-exp1}
B.~B.~Mandelbrot, D.~E.~Passoja, and A.~J.~Paullay,
Nature {\bf 308}, 721 (1984).

\bibitem{eta-exp2}
F.~C\'{e}lari\'{e}, S.~Prades, D.~Bonamy, L.~Ferrero,
E.~Bouchaud, C.~Guillot, and C.~Marli\`{e}re, 
Phys. Rev. Lett. {\bf 90}, 075504 (2003).

\bibitem{branch-exp}    
E.~Sharon and J.~Fineberg, 
Nature {\bf 397}, 333 (1999).

%Phase field
\bibitem{branch-phf}
A.~Karma and A.~E.~Lobkovsky,
Phys. Rev. Lett. {\bf 92}, 245510 (2004).

\bibitem{branch-mc2}
A.~Parisi and R.~C.~Ball, e-print, cond-mat/0403638.

\bibitem{gel}
 Y.~Tanaka, K.~Fukao, Y.~Miyamoto, and K.~Sekimoto,
Eurohpys. Lett. {\bf 43}, 664 (1998).

\bibitem{voro}
D.~Stoyan, W.~S.~Kendall, and J.~Mecke,
{\it Stochastic Geometry and its Applications}
(Wiley, Chichester, 1995).

%%%\bibitem{branch-th} branch-th

\bibitem{fuse}
G.~G.~Batrouni and A.~Hansen,
Phys. Rev. Lett. {\bf 80}, 325 (1998).

\bibitem{spring}
M.~Sahimi and S.~Arbabi,
Phys. Rev. Lett. {\bf 77}, 3689 (1996).

\bibitem{landau}
L.~D.~Landau and E.~M.~Lifshitz,
{\it Theory of Elasticity},
(Butterworth-Heinemanni, London, 1995).

\bibitem{ig-mf}
D.~G.~Harlow, H.~M.~Lu, J.~A.~Hittinger, T.~J.~Delph and R.~P.~Wei,
Modelling Simul. Mater. Sci. Eng. {\bf 4}, 261 (1996).

\bibitem{aniso}
%%fluctuation on anisotropy
M.~S.~Wu and J.~Guo,
J. App. Mech. {\bf 67}, 50 (2000).

\end{thebibliography}
\end{document}